\documentclass[fleqn,useAMS,usenatbib]{mnras}
\usepackage{graphicx}
\usepackage{latexsym,amssymb}
\usepackage{amsmath,morefloats}
\usepackage{natbib,graphicx,amsmath,color}
\usepackage{lmodern}
\graphicspath{{figures/}}
\newcommand{\vecM}{\mbox{\boldmath $M$}}
\newcommand{\vecg}{\mbox{\boldmath $g$}}

\newcommand{\veck}{\mbox{\boldmath $k$}}
\newcommand{\vecQ}{\mbox{\boldmath $Q$}}

\newcommand{\vecD}{\mbox{\boldmath $D$}}
\newcommand{\matR}{\mbox{$\bf R$}}

\newcommand{\bnabg}{\boldsymbol{\nabla_g}}
\newcommand{\galsim}{\texttt{GALSIM}}
\newcommand{\ngmix}{\texttt{ngmix}}
\newcommand{\nnsim}{\texttt{nsim}}
\newcommand{\snr}{$S/N$}

\newcommand{\coadd}{{\rm coadd}}

\newcommand{\model}{\mbox{$\boldsymbol{m}$}}
\newcommand{\modelc}{\mbox{$\boldsymbol{m}_\coadd$}}

\newcommand{\mcal}{\textsc{metacalibration}}
\newcommand{\Mdet}{\textsc{Metadetection}}

\newcommand{\Mcal}{\textsc{Metacalibration}}
\newcommand{\vest}{\mbox{\boldmath $e$}}
\newcommand{\est}{e}
\newcommand{\mcalR}{\mbox{\boldmath $R$}}
\newcommand{\mcalRS}{\mbox{\boldmath $R_S$}}
\newcommand{\gest}{\mbox{\boldmath $\hat \gamma$}}

\newcommand\edit[1]{\textcolor{black}{#1}}
\title[The little coadd that could]{The little coadd that could:  Estimating shear from coadded images}

\author[Armstrong et~al.]{Robert Armstrong$^1$,$^2$\thanks{\tt armstrong46@llnl.gov}, Erin Sheldon$^3$, 
  Eric Huff$^4$, 
  Jim Bosch$^2$,
  Eli Rykoff$^{5,6}$,
  \newauthor
  Rachel Mandelbaum$^7$, 
  Arun Kannawadi$^{2,8}$,
  Peter Melchior$^{2,9}$,
  Robert Lupton$^2$,
  \newauthor
  Matthew R. Becker$^{10}$,
  Yusra Al-Sayyed$^2$,
  The LSST Dark Energy Science Collaboration
  \\$^1$Lawrence Livermore National Laboratory, Livermore, CA 94551, USA
  \\$^2$\edit{Department of Astrophysical Sciences, Princeton University, Princeton, NJ 08544, USA}
  \\$^3$Brookhaven National Laboratory, Bldg. 510, Upton, NY 11973, USA 
  \\$^4$Jet Propulsion Laboratory, California Institute of Technology, 4800 Oak Grove Dr., Pasadena, CA 91109, USA
  \\$^5$Kavli Institute for Particle Astrophysics \& Cosmology, P. O. Box 2450, Stanford University, Stanford, CA 94305, USA
  \\$^6$SLAC National Accelerator Laboratory, Menlo Park, CA 94025, USA
  \\$^7$McWilliams Center for Cosmology and Astrophysics, Department of Physics, Carnegie Mellon University, Pittsburgh, PA
  \\$^8$\edit{Department of Physics, Duke University, Durham, NC 27708, USA}
  \\$^9$Center for Statistics and Machine Learning, Princeton University, Princeton, NJ 08544, USA
   15213, USA
  \\$^{10}$ High Energy Physics Division, Argonne National Laboratory, Lemont, IL 60439, USA
}

\begin{document}
\maketitle

\begin{abstract}
Upcoming wide field surveys will have many overlapping epochs of the same 
region of sky.  The conventional wisdom is that in order to reduce the 
errors sufficiently for systematics-limited measurements, like 
weak lensing, we must do simultaneous fitting of all the epochs.  Using current 
algorithms this will require a significant amount of computing time and effort. 
In this paper, we \edit{revisit the potential of using coadds for shear measurements.}  
\edit{We show on a set of image simulations that the multiplicative shear}
bias can be constrained below the 0.1$\%$ level on coadds, which is sufficient
for future lensing surveys.
We see no \edit{significant differences between simultaneous fitting and coadded approaches} for
two independent shear codes: \Mcal\ and BFD.  One 
caveat of our approach is the assumption of a principled
coadd, i.e.\ the PSF is mathematically well-defined for all the input images.
This requires us to reject CCD images that do not fully cover the coadd region.  We estimate 
that the number of epochs that must be rejected for a survey like LSST is on the 
order of \edit{$\sim 20\%$, resulting in a small loss in depth of less than $\sim 0.1$ magnitudes.
We also put forward a cell-based coaddition scheme that meets the above requirements for unbiased weak lensing shear estimation in the context of LSST.}
\end{abstract}

\section{Introduction}

Future large scale astronomical surveys such as \edit{the Vera C.\ Rubin Observatory 
Legacy Survey of Space and Time (LSST)}  \citep{LSST2009,2019ApJ...873..111I}, Euclid 
\citep{Laureijs2011}, and \edit{the Nancy Grace Roman Space Telescope (Roman)} \citep{Spergel2015} 
\edit{will} enable significant advances in many astronomical fields from the solar system to large 
scale cosmology through their unprecedented increase in the combination of depth and area coverage.
The observing strategy for these surveys will repeatedly 
\edit{image} the same patch of sky in different filters.  The number of observations 
varies significantly between surveys and can range from only a handful to 
hundreds of exposures.  Having repeated visits helps to build up the required 
depth, fill-in chip gaps, \edit{reject artifacts like cosmic rays and help achieve Nyquist sampling}.  Multiple 
observations can also help reduce systematic uncertainties, especially if they 
are taken on different parts of the camera and with different camera orientations.

One of the most powerful cosmological measurements from these surveys involves 
measuring the small distortion of galaxy images due to intervening matter.  
Photons from distant galaxies will be deflected by massive structures along the \edit{line of} sight 
causing a change in the shape, brightness and position.  This is known as gravitational 
lensing.  When the changes to the galaxies are small \edit{it is referred to} as \edit{``weak''} lensing.  
Weak lensing causes the shape of nearby galaxies to be correlated as \edit{their photons} pass by the 
same structures.  Computing the correlation of the shape of galaxies at different angular separations 
has shown to be a powerful probe of the expansion history and growth of structure in the 
universe.  Many past and \edit{ongoing} surveys have utilized weak lensing \edit{by large-scale structure, "cosmic shear",} 
to constrain cosmological parameters (see \cite{Giblin2021,Amon2022, Seco2022,Dalal2023,Li2023b, des_kids} 
for recent results).  However, for future surveys, \edit{one of the main challenges} will be to ensure 
that the systematic errors are under control.  There are many systematic errors that are 
important for weak lensing including\edit{:} the choice of shear measurement algorithm, 
characterization of the \edit{point spread function (PSF)}, photometric redshifts and theoretical predictions. 
In this paper\edit{, we will focus on how to combine images from multiple 
observations of a ground-based survey, }where the images are Nyquist sampled.

Weak lensing surveys have taken approaches that can be classified 
into three categories:
\begin{enumerate}
\item \label{method:first} Resample all of the images to a common pixel grid and 
take a weighted sum to construct a single coadded image.  Measurements are 
performed on the coadd with no need to return to the individual images.
\item \label{method:second} Perform measurements on each epoch separately.  
Combine these measurements at the catalog level by taking a weighted average.  
\item \label{method:third} Create the coadd as in \ref{method:first}, but only 
use it to detect objects.  With this information, go back to each of the 
individual epochs and do a joint simultaneous measurement of all the images.
\end{enumerate}

The simplest option is \ref{method:first}, as it significantly reduces the 
complexity and computation time.  For a survey with many epochs this is a huge 
gain.  Many lensing surveys have chosen to work on coadded images 
\citep{Hoekstra2006, Leauthaud2007, Hettersheidt2007, Lin2012,Melchior2015, 
Jee2016, Dalal2023}.  While this approach is attractive, it can also induce 
a number of challenges.  First, we must ensure that the 
individual epochs are properly registered relative to another.  Mis-registered 
images can severely distort the shape of objects. Also, the PSF can be 
challenging to model on the coadd due to the discontinuous ``jumps'' in the 
shape and size of the PSF when crossing the boundary between detectors.  
In addition, when combining images where the size 
of the PSF varies widely, there will be loss in S/N by doing a simple average.  
We discuss these problems in more detail in Section~\ref{Section:Coaddition}.
Method \ref{method:second} has been used in fewer cases \citep{Jarvis2003}.  
The main problem with this approach is that there will be objects that do not 
have a large enough $S/N$ to be detected on any individual epoch, but will be 
detected on the coadd.  However, there has been some work recently on how one 
might do this \citep{Budavari2017}.

With the need to reduce systematic errors, a number of recent surveys have 
followed the hybrid method \ref{method:third}.  The \edit{Canadian-France-Hawii Telescope Lensing Survey} 
\citep[CFHTLens;][]{Heymans2013} and \edit{Kilo-Degree Survey} \citep[KiDS;][]{Giblin2021} 
collaborations performed photometry and detection on the coadds and shear measurements via multi-epoch 
fitting. For the Dark Energy Survey (DES; \citealt{Seco2022}), only object 
detection was done on the coadd and all other measurements were performed simultaneously
on individual epochs.  This approach has worked well to reduce the systematic 
errors sufficiently low for cosmological analyses.  Can this multi-epoch 
fitting approach be applied to future surveys as well?
As the number of exposures increases, the computational demand scales as well.  
Existing surveys typically have fewer than $\sim10$ epochs to fit 
simultaneously. \edit{For the Hyper-Suprime Cam Survey (HSC), a precursor to LSST, 
model fitting on coadds is already $\sim40\%$ of the compute time. Extending this 
to work on multiple exposures will severely strain the computing budget when the number of exposures is large.} Again, being able to instead work on coadded images could 
significantly reduce the processing burden.  There may also be other 
reasons to prefer working on coadds.  For example, when separating blended objects into 
their individual sources, there is limited information available on 
\edit{any single epoch image, making method \ref{method:second} far from optimal.
The coadd is a conveniently high $S/N$ image, reducing a potentially large bookkeeping 
problem with many single epoch images when using method \ref{method:third}.}  

In this paper, we explore the possibility of avoiding multi-epoch fitting by 
demonstrating that we can recover shear on coadds \edit{at the level required for future surveys, particularly LSST}. 
\edit{In Section~\ref{Section:Coaddition}, we discuss the problems using 
coadds and  and potential ways to mitigate them.    
Section~\ref{Section:Simulation} describes a set of simulations to test
shear bias on both multi-epoch and coadd data.
We test two separate shear measurement pipelines: \edit{Bayesian Fourier Domain (BFD)} and 
\Mcal.  We summarize how these algorithms were configured and applied 
in Section~\ref{Section:Measurement}.  Section~\ref{Section:Results} shows the 
results of these methods.  We show that for both shear methods, we are able to 
recover the input shear signal with sufficient accuracy for both methods.     
Section~\ref{Section:Exclusion} examines the loss in $S/N$ from the 
coaddition itself and from rejecting exposures that would introduce
a PSF discontinuity from their edges in LSST simulations.
Finally, section \ref{Section:Cell} proposes a strategy to build
edge-free coadds in small regions called "cells" in the context of LSST data processing.}

\section{Coaddition}
\label{Section:Coaddition}

The simplest means of combining images is to use a direct weighted mean, 
median or clipped mean\edit{, with each image having a single weight}.  
The median and clipped mean are frequently used 
to reject artifacts in an image.  None of these are optimal in 
\edit{measuring the flux, size or signal-to-noise ratio ($S/N$) because 
they do not properly account for variations in PSF sizes and background levels. 
A more important issue, however, is that not all of these methods result in the coadd image having a 
well-defined PSF, by which we mean that the final
coadded image is the convolution of the true image with a coadd PSF at every point.
}

\edit{
As shown in~\cite{Mandelbaum2023}, there are strict
requirements that must be met in order for a coadd image to have a valid PSF.
The first requirement is that the coaddition scheme must be linear in the
pixel values.  They formally demonstrate that all non-linear coaddition procedures 
fail to to produce a well-defined PSF. For example, in a median or clipped coadd
the cores of brighter stars end up being clipped from the 
best seeing images, resulting in a flux-dependent (i.e., ill-defined) PSF.}

\edit{
~\cite{Mandelbaum2023} showed that a linear coaddition scheme must also have
either (a) all input images with the same PSF or (b) weights that are independent
of the true signal. Using the Poisson noise of the signal in the
weights will result in an ill-defined PSF. One must also be careful
of using spatially-varying weights for the input images as they can introduce 
additional scatter in the coadd.} Therefore, the simplest method to 
construct a \edit{coadd with a} valid PSF is to use a weighted mean where the weight only 
depends on the background.

\edit{There are more sophisticated approaches such as \cite{Rowe2011} which constructs
the coadd PSF as a linear combination of pixel values of the individual exposures in such 
a way to produce a well sampled coadd from undersampled input images. This will be important
for space-based lensing surveys for which most images are not Nyquist-sampled. There has been 
renewed interest from the Roman science team in using this approach for
shear measurements \citep{Hirata2023,Yamamoto2023}}

For the majority of detections, which are small faint objects, the epochs with the 
smallest seeing will have the most information.  Under certain assumptions, 
\cite{Kaiser2004} and \cite{Zackay2017} constructed an optimal coadd in terms of both PSF 
size and information content.  This results in an image that loses no information when 
combining epochs of different seeing.  Further research in this area seems promising in order to 
maximize the \edit{measured} $S/N$ for a coadd.  However, for this paper we only consider the 
scenario of using a simple weighted mean.  

\subsection{Image Registration and Noise}
\label{Subsection:Registration}
\begin{figure*}
\includegraphics[width=1\textwidth]{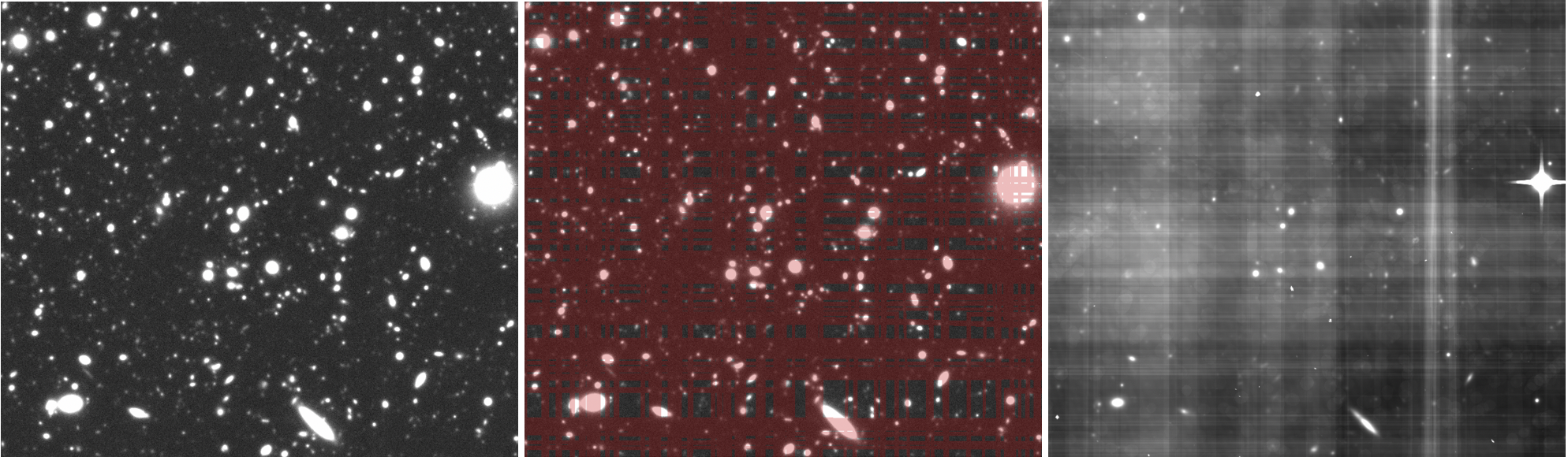}
\caption{
\edit{Left: A coadded image from the HSC UltraDeep field taken in the i-band from $\sim$200 overlapping images.  Middle: The same image as on the left with red lines showing the edges of CCDs.  It can be seen that almost every object overlaps an edge.  Right: The variance map of the image where darker colors indicate lower variance. The lattice structure is due to the small dither patterns for the HSC UltraDeep field.}}
\label{fig:noise}
\end{figure*}

In order to coadd, we must account for the actual pointing of each image and 
the local distortion caused by the optics, atmosphere, etc.  This requires 
resampling and interpolating each image onto the same coordinate grid.  This 
interpolation \edit{procedure introduces noise correlations from one pixel to another.} 
In addition, depending on the local distortion and relative offset 
between pixel grids, this can cause the noise to become non-stationary, i.e. it 
can vary across the image.  Figure~\ref{fig:noise} shows \edit{a coadded image
 and variance map from the HSC UltraDeep survey \citep{HSC_DR1} where there are $\sim200$ $i$-band images
 that were used to create the coadd.  The image cutout corresponds to 
a region roughly $2\arcmin \times 2\arcmin$ in area.  In the variance map, the darker regions have 
more observations and hence lower variance.  The lattice-like structure corresponds to CCD edges being shifted
due to the small dithering strategy employed in the UltraDeep region.
The noise in this image is dominated by the changing number of input images, but smaller variations
can be seen due to resampling and grid offsets.}  The range of variations is 
$\sim10\%$ and the scale on which they vary is close to the average size of on\edit{e} 
object (see~\ref{Section:Exclusion}).

By definition, the variance map does not include the pixel to pixel covariance. 
This will also vary spatially depending on the dither pattern and 
interpolation kernel.  For typical observing patterns in HSC, the amount of 
variance lost to covariance is $\sim 10\%$.  The correlated noise pattern must be 
taken into account to achieve the precision in weak lensing. \cite{Gurvich2016} 
showed that ignoring such correlations results in errors of a few percent, well 
above the requirements of future surveys.  \edit{However, when such correlations are known 
and accounted for, their impact can be mitigated.}

With a complete knowledge of the astrometric shifts and distortions, it is, in 
principle, possible to construct the full pixel covariance for a given region.  
However this is far from practical and few \edit{existing algorithms} would be able 
to take advantage of such information.  As an alternative, we follow \cite{SheldonHuff2017} and take 
Monte Carlo realizations of the noise for each epoch and coadd them in the same 
way as the images. \edit{This produces a single noise image for each coadd image that
includes the pixel correlations induced from interpolation. This noise image will be utilized when computing 
the shear in our image simulations. For our tests in this paper, we assume that the noise in each exposure is Gaussian.  
We found that a single noise realization per epoch was
sufficient to characterize the noise.} 

\subsection{\edit{Edge-Free Reconstruction}}
\label{Subsection:PSF}
In many situations, we need to be able to model the PSF from the observed stars 
and interpolate it to the positions of galaxies.  Typically, a low order 
polynomial is used to describe the variation of the PSF across a CCD.  The 
challenge on a coadd is that the PSF is discontinuous across the edge 
of a detector from any individual epoch.  Given this, it can be difficult to construct a 
model on the coadd.  An alternative approach used by 
\cite{Jee2016} and \cite{Bosch2018} is to create a valid coadded PSF by resampling 
and interpolating the single epoch PSFs in the same way as the images. This avoids any problems 
of having to spatially interpolate a complex PSF model.  However, objects 
that include an epoch which crosses a detector boundary may pose a problem, as the PSF is not well defined.  
For HSC DR1, objects that contained such a boundary image were flagged and 
corresponded to $\sim15\%$ of the detected objects.  Because the PSF is 
potentially inaccurate, these objects \edit{may} need to be removed from scientific 
analyses \edit{that need an accurate PSF}.  \edit{Simply removing such} objects will not 
be a viable option for a survey like LSST, as almost 
every object will cross at least one CCD boundary.  Instead, we propose to 
simply throw out epochs that cross a detector boundary when \edit{creating the coadd image, PSF and noise
image.}   The number of rejected images will depend on the size of the coadd, with fewer rejections for smaller coadds.  We discuss the loss of depth from rejecting these visits in 
Section~\ref{Section:Exclusion}.

\section{Simulations}
\label{Section:Simulation}
\edit{In order test our ability to recover shear on edge-free coadds,} we construct a suite of simulations to test the impact of coaddition specifically for shear estimators.  The goal is to 
simulate multiple epochs of the same galaxy with a known shear and compare the 
inferred shear from both multi-epoch and coadd measurements.
The observing conditions for each epoch (noise, PSF, etc.) are 
simulated to roughly match observations from DES data.  While DES will have 
only ten epochs by the end of the survey, it still provides a useful test while 
reducing somewhat the computation time.  Even with 10 epochs, the computation 
time to generate the dataset is significant as we need to simulate billions of 
galaxies to reach the needed precision for cosmic shear.  With fewer epochs, we 
may also be more sensitive to certain systematic errors that could get averaged 
down for a survey with many more observations.
To generate the galaxy population we use code from the \nnsim\ 
package\footnote{\url{https://github.com/esheldon/nsim}} which is based on \galsim\ 
and \ngmix.  The generated galaxy images have the following properties:

\begin{itemize}
\item[--] We generate single galaxy images on $48 \times 48$ postage stamps with no neighbors. 
          The pixel scale is the same as DES, 0.27 arcseconds/pixel.  The pixel noise is uniform
          for each visit and is modeled as a zero-mean Gaussian \edit{with $\sigma$ chosen to be typical of DES $i$-band images}. 
\item[--] We assume the PSF model is perfectly known.  The model is a Moffat \edit{profile \citep{Moffat1969}} with $\beta=3.5$.
          The \textsf{FWHM} of the PSF varies for each epoch as a LogNormal distribution with mean 
          0.95 \edit{arcseconds} and $\sigma_\text{FWHM}=0.1$.  Values are truncated between 0.8 and 1.15.  The PSF ellipticity is
          Gaussian \edit{distributed} with a mean of 0.0 and 0.01 for $e_1$ and $e_2$ respectively, and $\sigma_e=0.01$.
          We use the same offset in PSF image as for the galaxy in its image, so that the
          same shifts must be applied when resampling (see \S \ref{Section:Results}).
\item[--] The galaxy models are pure exponentials.  We sample from an empirical model for the flux 
          and size based on \textsf{COSMOS} data.  We used the \textsf{COSMOS} I$<$25.2 dataset packaged 
          along with the \galsim\ software to construct a kernel density estimator for the size and flux.  
          For the \edit{bandwidth} of the estimator we use a kernel with 0.1$\times$ the intrinsic data covariance.  
          The intrinsic ellipticity is drawn from the distribution $P(e) \propto (1-e^2)^2 \exp{(\edit{-}e^2/2\sigma_e^2)}$ \citep{Bernstein2016}
          with $\sigma_e=0.02$.  \edit{In the real Universe $\sigma_e$ is much larger, $\sim0.3$. The small width chosen 
          here is set to reduce the computing time}.  All galaxies are sheared by $g_1=0.02$, \edit{where $g$ is the reduced shear and $g_1$ is the component parallel to the $x$ and $y$ axes.}
\item[--] We generate 10 epochs for each object.  Each visit has a random shift applied \edit{in $x$ and $y$ that is uniform over a pixel.}  The Jacobian values of the WCS for each exposure are allowed to vary for each epoch.  
          The amount of variation is taken from values measured on DES.  For coaddition, we compute a mean weighted by the inverse of the pixel variance. To resample the images we use a 3rd order Lanczos kernel.  
\item[--] We construct a coadded noise image, as described in Section~\ref{Subsection:Registration}, by 
          generating a single Monte Carlo realization of the noise for each epoch and processing it with the same resampling used for the image, and summed with the same weights.  The coadded PSF is constructed in a similar way by combining the individual PSFs for each epoch.

\end{itemize}

The choice of using simplified galaxy models was employed solely to reduce the 
computational run time.  For more complicated models there was significant 
computation time in generating the images.  
\edit{Even with the simplifications listed, the computational cost 
is on the order of 10 million CPU-hours. }We did some limited testing of more 
complicated galaxy models with reduced statistics, and these showed no sign of 
problems.  Since both of the shear measurement pipelines employed here have 
already shown to be largely insensitive to the type of galaxy model chosen, we 
do not expect this to affect our results.

\section{\edit{Shear} Measurement Methods}
\label{Section:Measurement}

\edit{There are a number of shear measurement methods that have demonstrated the 
potential to meet the strict requirements for upcoming surveys. 
These include model fitting approaches
\citep{Miller2013}, self-calibration methods \citep{Li2023a}, and Bayesian 
forward modeling \citep{Schneider2015}. To test our approach to measuring
shear on coadds, we select the \Mcal\ and BFD algorithms. Both of these 
methods have the capability to measure shear on coadd and multi-epoch data. 
LSST has adopted \mcal as an official data product and the addition BFD as
an additional method gives us confidence that other methods can similarly 
be adapted to work on coadd data. We briefly 
summarize the \Mcal\ and BFD methods and describe how they were applied \edit{to} these 
simulations.}

\subsection{BFD}
\label{Section:BFD}

\edit{
The Bayesian Fourier Domain (BFD) method \citep{Bernstein2014,Bernstein2016} 
was designed to overcome some of the inherent biases that arise from the 
measurement of galaxy shapes. It does not measure the shapes of
individual galaxies, but rather takes a probabilistic approach, estimating
the probability that a galaxy would produce the pixel data for a given shear.
If we assume that a set of galaxies is sheared by
a constant value $\vecg$, we can write the shear posterior probability by Bayes theorem as:
\begin{equation}
    P(\vecg | \vecD) = \frac{P(\vecD | \vecg) P(\vecg)}{P(\vecD)},
\end{equation}
where $P(\vecD|\vecg)=\Pi_i P(D_i|\vecg)$ is the probability of obtaining the pixel 
data $D_i$ for the whole galaxy population.  
For a single galaxy we can write
\begin{equation}
P(D_i|\vecg) = \int d\vecM_{\rm true}P(\vecM_{\rm obs}|\vecM_{\rm true})P(\vecM_{\rm true}|\vecg),
\end{equation}
where we have marginalized over a set of true measurements $\vecM_{\rm true}$.  
}
\edit{
If we assume that
the lensing is weak, we can do a Taylor expansion around $\vecg=0$
\begin{align}
P(D_i | \vecg) & = P_i + \vecQ_i\cdot \vecg + \frac{1}{2} \vecg \cdot
\matR \cdot \vecg, \\
P_i & \equiv P(D_i| \vecg=0)\nonumber\\
\label{q1eqn}
\vecQ_i & \equiv \left.\bnabg P(D_i | \vecg)\right|_{\vecg=0} \nonumber\\
\matR_i & \equiv \left. \bnabg \bnabg P(D_i | \vecg)\right|_{\vecg=0}. \nonumber
\end{align}
The $P_i$ values are the probability of measuring $\vecM_{\rm obs}$ given the noise-free intrinsic distribution $\vecM_{\rm true}$.  We can use observations of a deeper dataset to approximate the true distribution.  Most ongoing and future surveys contain a dataset suitable for this purpose.  The $\vecQ_i$ and $\matR_i$ give the differential probability of observing a set of measurements under shear. If we assume that the likelihood is independent of shear, the derivatives propagate to the intrinsic distribution $P(\vecM_{\rm true}|g)$. For the measurements we choose below and a fixed weight function, these values can be computed analytically.
}

\edit{
We can now rewrite the shear posterior over many sources as
\begin{align}
-\ln P(\vecg | \vecD) & = (\rm const) - \ln P(\vecg) - \sum_i \ln
                        P(\vecD_i | \vecg) \\
 & =  (\rm const) - \ln P(\vecg) - \vecg \cdot \vecQ_{\rm tot}
+ \frac{1}{2} \vecg \cdot \matR_{\rm tot}
   \cdot \vecg, \nonumber \\
\label{qtot}
\vecQ_{\rm tot} & \equiv \sum_i   \frac{\vecQ_i}{P_i} \\
\label{rtot}
\matR_{\rm tot} & \equiv \sum_i \left(
   \frac{\vecQ_i\vecQ_i^T}{P_i^2} - \frac{\matR_i}{P_i}\right)
\end{align}}
\edit{If we ignore the prior, we can write the shear posterior distribution as a Gaussian in \vecg,
with mean value
\begin{equation}
\label{gbar}
\bar\vecg = \matR_{\rm tot}^{-1} \vecQ_{\rm tot}.
\end{equation}}
\edit{We can account for selection effects by adding an additional term to both $\vecQ_{\rm tot}$ and $\matR_{\rm tot}$ that computes the probability we would select an object given a set of cuts on our measurements. It is important that the selection criteria only depend on our set of measurements $\vecM_{\rm obs}$. We use the postage stamp correction scheme as described in Section~2.3 of \cite{Bernstein2016}.}

\edit{
For our measurements we compress the pixel data into a set of moments, $\vecM_{\rm obs}$ in Fourier space
\begin{equation}
\label{moments}
\vecM_{\rm obs} \equiv \left( \begin{array}{c}
M_f \\
M_r \\
M_+ \\
M_\times
\end{array}
\right) = \int d^2k\, \frac{\tilde I^o}{\tilde T(\veck)}
W(|\veck^2|) 
\left( \begin{array}{c}
1 \\
k_x^2 + k_y^2 \\
k_x^2 - k_y^2 \\
2 k_x k_y
\end{array}
\right),
\end{equation}
where $\tilde I^o$ is the Fourier transform of the image and $\tilde T(\veck)$ is the Fourier 
transform of the PSF and $W(|\veck^2|)$ is a fixed weight function to bound the noise. 
}

As noted in \cite{Bernstein2016}, because the BFD moments are linear, 
measurements on multiple exposures can be combined at the catalog level by 
taking weighted sums of the individual moments.  This is true as long as each 
individual exposure is unaliased.  This greatly simplifies doing a 
multi-exposure measurement compared to other methods because there is no need to 
go back to the individual pixels.  
\edit{For measurements on the} coadd, we need to have an accurate measure of the pixel noise power 
spectrum.  We can measure this directly from the resampled and coadded noise images.

We configure the BFD code to run with the following parameters:
\begin{itemize}
\item We use the `$k\sigma$' weight function defined in \cite{Bernstein2016}:
\edit{
\begin{equation}
\label{ksigma}
W\left(|k^2|\right)  \equiv \left\{ 
\begin{array}{cc}
\left( 1 - \frac{k^2\sigma_k^2}{2N}\right)^N & k <
                                             \frac{\sqrt{2N}}{\sigma_k} \\
0 & k \ge
                                             \frac{\sqrt{2N}}{\sigma_k} 
\end{array}
\right.
\end{equation}
}
The parameters are set to $N=4$ and $\sigma_k=1$.  We chose 
$\sigma_k$ to be \edit{ smaller than in \cite{Bernstein2016}. This removes the small shear bias observed 
in the image simulations from that study.} Other parameters were chosen to match those listed in the 
\galsim\ settings from Table 1 in \cite{Bernstein2016}.
\item We simulated a deep dataset with the same settings, except the pixel 
noise was reduced by a factor of 100.
\item We divide the data into two different regimes: flux signal-to-noise ($S/N$) \edit{between} 5-25 and \edit{between} 25-50.  We 
build a separate prior for each regime.  For the higher $S/N$ range, we doubled 
measured moment noise to increase the overlap with the relatively small number of deep galaxies.
This is to account for the fact that higher $S/N$ galaxies will have a smaller
measurement error and narrower likelihood, and will therefore overlap fewer templates.  See \cite{Bernstein2016}
Section 2.5 for more details.
\end{itemize}

\subsection{\Mcal}
\label{Section:Metacal}

\Mcal\ is a method to calibrate weak lensing shear measurements using
transformations of the real images, without reference to external data sets or
simulations \citep{HuffMand2017,SheldonHuff2017}.  Below we give a brief review
of the \mcal\ formalism.  For full details of the \Mcal\ implementation used in this work, see
\cite{SheldonHuff2017}.

The basic assumption behind \mcal\ is that the two-component shear estimator \vest\
is linearly related to the applied shear $\vecg$, such that
\begin{align} \label{eq:Eexpand}
    \vest &= \vest|_{g=0} + \frac{ \partial \vest }{ \partial \vecg}\bigg|_{g=0} \vecg  + ... \nonumber \\
          &\equiv \vest|_{g=0} + \mbox{\mcalR}\vecg  + ...
\end{align}
We have defined the $2 \times 2$ shear response matrix
\begin{align}
    \mbox{\mcalR} &\equiv \frac{\partial \vest}{\partial \vecg} \bigg|_{g=0}.
\end{align}
With \mcal\ these derivatives are performed by shearing the real image
of the object and calculating a finite difference derivative.  The image
is deconvolved \edit{with} the PSF, sheared by a small amount, and reconvolved by
a larger PSF to suppress amplified noise.  Measurements $\vest$ are then made
on these images and the derivatives are formed. For element $i,j$
\begin{equation} \label{eq:Rnum}
    R_{i,j} = \frac{\est_i^+ - \est_i^-}{\Delta g_j}, 
\end{equation}
\edit{where $\est^+$ is the measurement made on an image sheared by
$+g$, $\est^-$ is the measurement made on an image sheared by $-g$, and $\Delta g$ = 2$g$.}
For the measurement of simple mean shear, which is all we will  use in this
work, the calibrated estimator is formed using the ensemble mean of the
estimator and response
\begin{align}
    \langle \vest \rangle &= \langle \vest \rangle |_{g=0} + \langle \mbox{\mcalR} \vecg \rangle + ... \nonumber \\
                          &\approx \langle \mbox{\mcalR} \vecg \rangle,
\end{align}
and thus
\begin{align} \label{eq:rcorr}
    \langle \gest \rangle &= \langle \mbox{\mcalR} \rangle^{-1}  \langle \vest \rangle \approx \langle \mbox{\mcalR} \rangle^{-1} \langle \mcalR \vecg \rangle.
\end{align}

Note that if selections are performed on the measurements, an additional
response \mcalRS\ must be calculated and added to the ensemble response
to correct for shear-dependent selection effects \citep{SheldonHuff2017}.

The convolutions and shears applied to the images result in correlated noise
that can bias the shear estimate. We apply a simple empirical correction. We
generate a noise image with the same amplitude as the noise in the real data,
and the same shape as original image.  We rotate the noise image by 90 degrees.
We apply \mcal\ shearing and convolution operations.  We then rotate this field
back by 90 degrees and add it to the image of the object before performing a
measurement. This cancels the correlated noise bias in the ensemble shear
measurement \citep{SheldonHuff2017}.

Resampling involves interpolation of the images, which introduces additional
correlated noise.  This can be dealt with in the same correlated noise
correction scheme.  We generate noise images as described above, but pass them
through resampling and summing before passing them through the
\mcal\ convolution and shearing operations.

\section{Simulation Results}
\label{Section:Results}

We ran the BFD and \mcal\ shear measurement codes on
our simulations in both multi-epoch mode and on the coadded images. 
\edit{Each method was used on independent data sets so the resulting errors are uncorrelated.} 
For \mcal\ we made consistent cuts on coadd and multi-epoch \edit{objects such that
\snr$ > 10$ and square size relative to the PSF of $T/T_{\mathrm{PSF}} > 0.5$.  We used the size defined as $T=\langle x^2 + y^2\rangle$ 
from the second moments of the brightness profile.}
For BFD we made cuts \edit{on the flux} of $5 < \snr < 50$ as described in \ref{Section:BFD}.
We applied corrections for selection effects in both methods \edit{which are needed to get the required precision}.  

\edit{We tested the accuracy of
our methods by comparing the measured value of $g_1$ to the input of the simulations.  One common
metric is to compute the multiplicative bias $m$ and additive bias $c$ defined by
\begin{equation}
g_{\rm meas} - g_{\rm true} = m g_{\rm true} + c.
\end{equation}
We found that $c$ was consistent with zero for all the tests we ran and therefore show only the results for $m$.
In Table~\ref{tab:results}
we summarize the measured bias on $g_1$ for each shear method and for coadd vs. multi-epoch fitting.}
\edit{For $g_2$, we found the measured values
consistent with zero as expected.}  These results show
that there is no additional bias in either shear method when measuring the
shear on the coadds.  \edit{Ambitious future surveys require the shear bias to be $<10^{-3}$ \citep{Huterer2006, Amara2008} in order to keep the bias from 
degrading the accuracy of cosmological analyses.  This number is for all weak lensing systematics that could
contribute to a bias in the shear.  Both methods
demonstrate that coaddition does not contribute the bias for these simulations.}

We note additional findings from our analysis: 
\begin{itemize}
\item The Monte Carlo noise images that we generated for each epoch were the
same size as the galaxy postage stamp; 48 by 48 pixels.  If the size of the
noise image is reduced we see a bias in the BFD results.  Reducing the noise
image to 40 by 40 results in a $\sim 1\%$ bias and increases with decreasing 
size.  Presumably, this comes from not having sufficient information to
construct the power spectrum accurately on the final coadd.  Note that \mcal\
has a strict requirement that the noise image must be at least as large as the
original image.
\edit{\item We placed the PSF from each image at the same sub-pixel offset as the
galaxy so that a shift was required during resampling.  This is required so that
the same small smoothing due to interpolation is present in both the image and PSF image.  
BFD did not seem sensitive to the centering of the PSF, while \mcal\ required a high
order interpolation kernel if the PSF was properly off center before resampling.  BFD was likely less sensitive
to this because the same centering error was in the deep reference set and the shallower images, 
so this problem was being calibrated out.}

\item We tested coadds created from the \edit{original simulation images that had large random
rotations as LSST will for camera rotations}.  We found no additional bias, as long as the images were padded
sufficiently to avoid corner effects.
\edit{\item A detailed study testing the accuracy of cell-based coadds (see Section~\ref{Section:Cell}) on LSST-like simulations has 
also been done for an extension of \Mcal\ called \Mdet\ \citep{Sheldon2023}. This study did more realistic 
simulations than this work including things like realistic galaxy models, galaxy blends, stars, and 
cosmic rays. They showed that \Mdet\ measured no shear bias to within the requirements of LSST.}


\end{itemize}

\begin{table}
\begin{tabular}{ll}
\hline
Sample & \edit{Multiplicative bias for $g_1$} \\ \hline
BFD Coadd & $0.0012 \pm 0.0007$ \\
BFD Multi-epoch & $-0.0014 \pm 0.0007$ \\
\mcal\ Coadd & $0.0007 \pm 0.0004$  \\
\mcal\ Multi-epoch & $0.0004 \pm 0.0004$  \\ \hline
\end{tabular}
\caption{The measured shear bias from simulations for BFD and \mcal\ on both 
coadd and multi-epoch data.  These results show no statistically significant 
bias for any of the methods beyond that expected from the breakdown
of the weak shear approximation ($\sim 0.0005$). \edit{Measured errors on the bias are 1$\sigma$}.}
\label{tab:results}
\end{table}

\section{\texorpdfstring{$S/N$}{} Loss in Edge-free Coadds}
\label{Section:Exclusion}
\edit{For edge-free coadds it
is important to understand how much information will be lost.  This lost information will
come from the coaddition process itself and also due to the rejection of edge exposures,
images that do not fully cover the coadd. While, some of this lost information could be recovered using optimal methods for coaddition, we again restrict our analysis to using a weighted mean for coaddition. }

\subsection{Expected Flux \texorpdfstring{$S/N$}{} Loss from Coaddition}
\label{Subsection:FluxSN}
For the simple weighted mean coadds used in this work, we expect measurements
on a coadd to be noisier than those using optimal multi-epoch fitting when the
size of the PSF varies between epochs.  We expect the increase in noise to be
largest for the smallest objects such as stars and distant galaxies, the images
of which are more affected by the PSF.

To gain some intuition, we derive the functional form of the expected loss in
$S/N$ for the case of a linear fit for the amplitude $A$ of a template, or
``matched filter''.  In this case we can analytically compute the uncertainty
in the estimated flux, $\hat{A}$, from a linear fit to the data (see Appendix
\ref{Section:FluxSNAppendix}). It is also useful to introduce a ``toy model'', which
illustrates the main features of this process.  For the toy model we use data
that follows a circular Gaussian, which makes the calculations easier and
allows introduction of useful approximations.

In Appendix~\ref{Section:FluxSNAppendix} 
we derive the full expression for the variance in the multi epoch and coadd
cases.  We also show that (see equation \ref{eq:vargal_appendix}), for the toy model and small
variations in the PSF size, the ratio of the flux variance derived from the
coadd to that from fitting the epochs simultaneously is approximately:
\begin{align} \label{eq:vargal}
\left( \frac{ {\rm var} \hat{A}_\coadd}{{\rm var} \hat{A}_{\rm multi} } 
\right)_{\rm toy} &
\approx 1 + 2 \left( 1 - R \right)^2 \left( \frac{\Delta 
\sigma_p}{\sigma_p} \right)^2,
\end{align}
where $\sigma_p$ is the mean PSF size, $\Delta \sigma_p$ is the RMS scatter in PSF
size across the individual visits, and $R=\sigma_g^2/(\sigma_p^2 + \sigma_g^2)$ 
is the ratio of the \edit{pre-seeing galaxy size $\sigma_g^2$ to the post-seeing galaxy size $\sigma_p^2 + \sigma_g^2$.}  This result shows 
the increased variance is proportional to $(1-R)^2$ and the relative variation in PSF size
$(\Delta \sigma_p/\sigma_p)^2$.  For a large galaxy $R$ is near unity and for a star
$R$ is near zero, so as expected the effect is largest for stars.

We tested the validity of equation~\eqref{eq:vargal} using a simple Monte Carlo
simulation.  We used the \galsim\ package \citep{GalSim} to generate images of \edit{circular} Gaussian
galaxies, convolved by \edit{circular} Gaussian PSFs, and used the \ngmix\ 
package\footnote{\url{https://github.com/esheldon/ngmix}} to fit for the flux. We used
a PSF with FWHM=0.9 \edit{arcseconds and pixel scale of 0.26}, with Gaussian variation between images of $\Delta
\sigma_p/\sigma_p = 0.1$.

\begin{figure}
\includegraphics[width=0.45\textwidth]{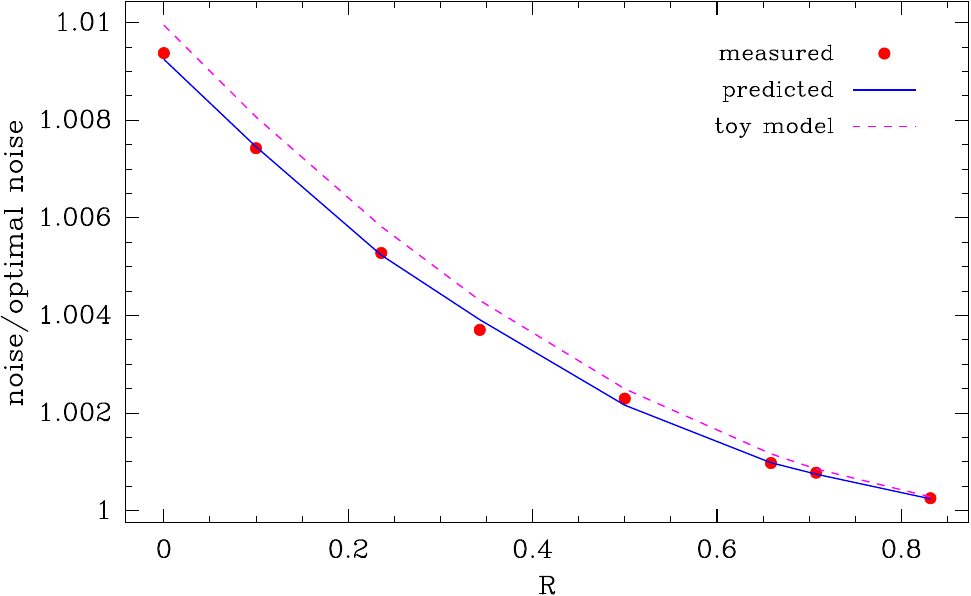}
\caption{The ratio of the noise in the estimated flux when measured on a coadd 
compared to an optimal combination of images when both the galaxy and PSF are Gaussian.
The horizontal axis is a dimensionless measure of the galaxy size compared to the total
PSF-convolved object size, $R=\sigma_g^2/(\sigma_p^2 +
\sigma_g^2)$, 
where $\sigma_g$ is the galaxy size and $\sigma_p$ is the PSF size.
The points are measured values from simulations, and the solid line is the 
\edit{ratio of equations~\eqref{eq:exact_var_coadd} and~\eqref{eq:exact_var_multi}}.
The dashed line shows 
the simple toy model presented in equation~\eqref{eq:vargal}. \label{fig:mcresults}}
\end{figure}

All objects were placed at the center of the image, so that no interpolation was 
required when coadding the images.  We used the same constant noise for all 
images, and used a simple mean for the coaddition process.  The primary 
variable of interest is the resolution $R$.  We varied the size of the galaxy 
from zero, or star-like, to more than twice the PSF size, such that $\langle R\rangle = 
0.83$.  A comparison of the results from the simulations and analytical formula 
are shown in Figure~\ref{fig:mcresults}.  The points are the measured values 
and the solid line is the ratio of the exact variance estimate from 
\eqref{eq:exact_var_coadd} and~\eqref{eq:exact_var_multi}.  The dotted lines are the 
predicted values from the toy model in equation~\eqref{eq:vargal}.  As expected
the exact estimators describe the results well.
The toy model slightly over-predicts the increase in variance, but is 
evidently useful for understanding how the effect scales with resolution. 
For a given survey we can use this to predict the effect that coaddition will 
have on the flux uncertainty. 

While the numbers used to generate Fig.~\ref{fig:mcresults} match the 
values we used in our simulations from Section~\ref{Section:Simulation}, surveys like LSST will 
have a wider range of seeing values and therefore larger $\Delta
\sigma_p/\sigma_p$. To estimate this value for LSST, we use data generated by the 
"Operations Simulator" ({\textsc{OpSim}})~\citep{2016SPIE.9910E..13D,2016SPIE.9911E..25R}. \textsc{OpSim} includes
an estimated seeing for all exposures over a full 10-year simulated LSST survey. Using a recent baseline observing schedule (version 3.2\footnote{\url{https://s3df.slac.stanford.edu/data/rubin/sim-data/sims_featureScheduler_runs3.2/baseline/baseline_v3.2_10yrs.db}}\citep{jones_r_lynne_2020_4048838}), 
we compute $\Delta\sigma_p/\sigma_p\approx 0.3$ and $\sigma_p \approx 0.4$ arcseconds.  
For these values, the ratio of the coadd noise relative to the optimal noise increases to $\sim$1.2 
for stars ($R=0$) and to $\sim$1.01 for the largest galaxies ($R=0.8$). For typical galaxies
used in weak lensing, this will not be a large effect, but may be a noticeable source of noise for smaller objects. One approach to
reduce the added noise would be to generate multiple coadds in different bins of seeing. This approach would reduce the number of simultaneous images to analyze and limit the range of seeing values.
We leave investigation of the optimal way to build LSST coadds for weak lensing to future work.

\subsection{\texorpdfstring{$S/N$}{} Loss from Rejecting Exposures}
We can estimate how many 
exposures would be rejected for different cell sizes 
\edit{using the previously mentioned \textsc{OpSim} data.  
It includes information on where the camera is pointing at any given time during the survey.
We use a slightly older \textsc{OpSim} strategy labeled as {\textsf{minion\_1016}} used by
the Dark Energy Science Collaboration to generate the Data Challenge 2 data~\citep{dc2}.}

\begin{figure}
\includegraphics[width=0.5\textwidth]{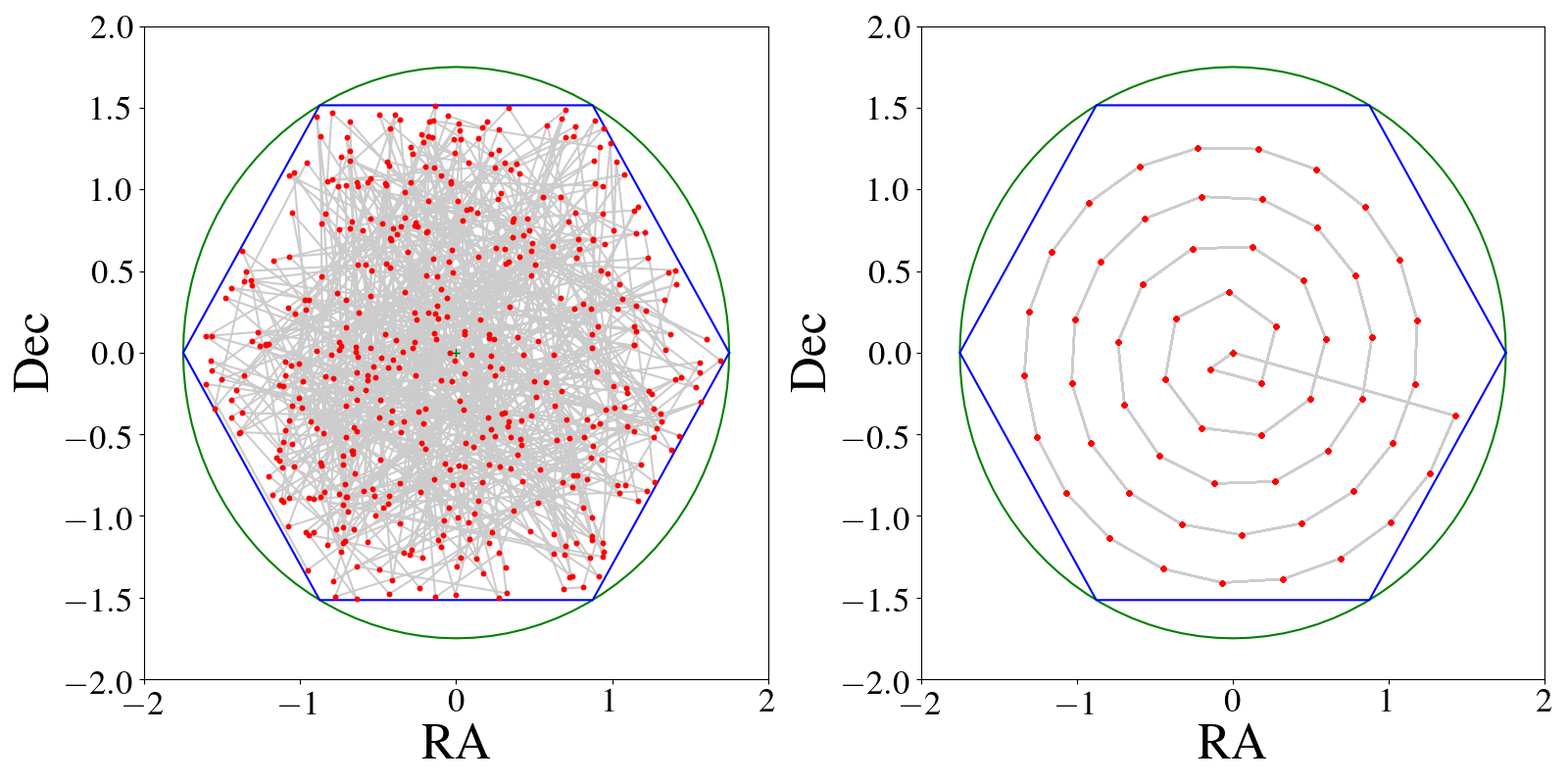}
\caption{
Two different LSST dither patterns are shown taken from \protect\cite{LSSTSims}.  
On the left each visit gets a random position and on the right a spiral pattern.
Each dot shows a separate offset from the nominal pointing.
The green circle represents the approximate size of the LSST focal plane.  
The dots are restricted to within the blue hexagon for this study. }
\label{fig:dither}
\end{figure}

By default the simulation does not provide a dithering strategy for 
different exposures of the same pointing, therefore we
we implemented a number of different dither strategies as provided by the LSST 
Metrics Analysis Framework (MAF) software \citep{LSSTSims}.  We found that the 
results had little dependence on which dithering strategy was chosen, therefore 
we will only show two representative dither patterns - a random offset for each 
visit and a spiral pattern. Fig.~\ref{fig:dither} shows the distribution of 
offsets for these two patterns.

\begin{figure*}
\includegraphics[width=\textwidth]{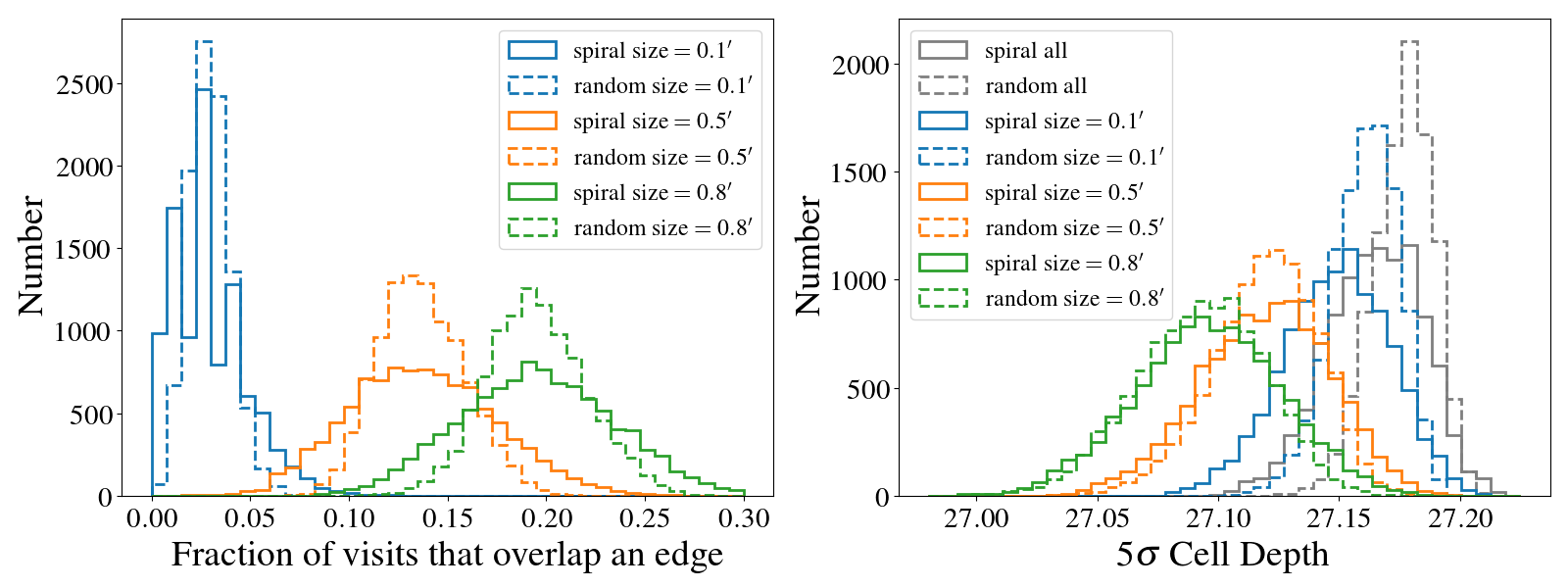}
\caption{
{\it Left}: The fraction of visits that would need to be rejected from a patch for three 
different patch sizes (0.1, 0.5 and 0.8 arcminutes) and the two dither patterns 
shown in Fig.~\ref{fig:dither}.  {\it Right}:  The corresponding distribution 
of 5$\sigma$ depths for each patch. }
\label{fig:cell}
\end{figure*}

We compute the number of missing exposures by placing 10,000 
square patches of a \edit{given} size on the sky and calculating the number of 
r-band visits that have a CCD edge that will cross each patch.  From this, we can calculate 
the reduced r-band 5$\sigma$ point source depth for each patch \edit{that would result from rejecting
these exposures}.  Fig.~\ref{fig:cell} 
shows \edit{the 5$\sigma$ cell depth} for three different cell sizes with side length 0.1, 0.5 
and 0.8 arcminutes. Given these cell sizes, the fraction of rejected visits 
can vary from 3$\%$ to 20$\%$.  (The variation in depth can also affect the 
science results, see \cite{Heydenreich2020})

\begin{figure}
\includegraphics[width=0.5\textwidth]{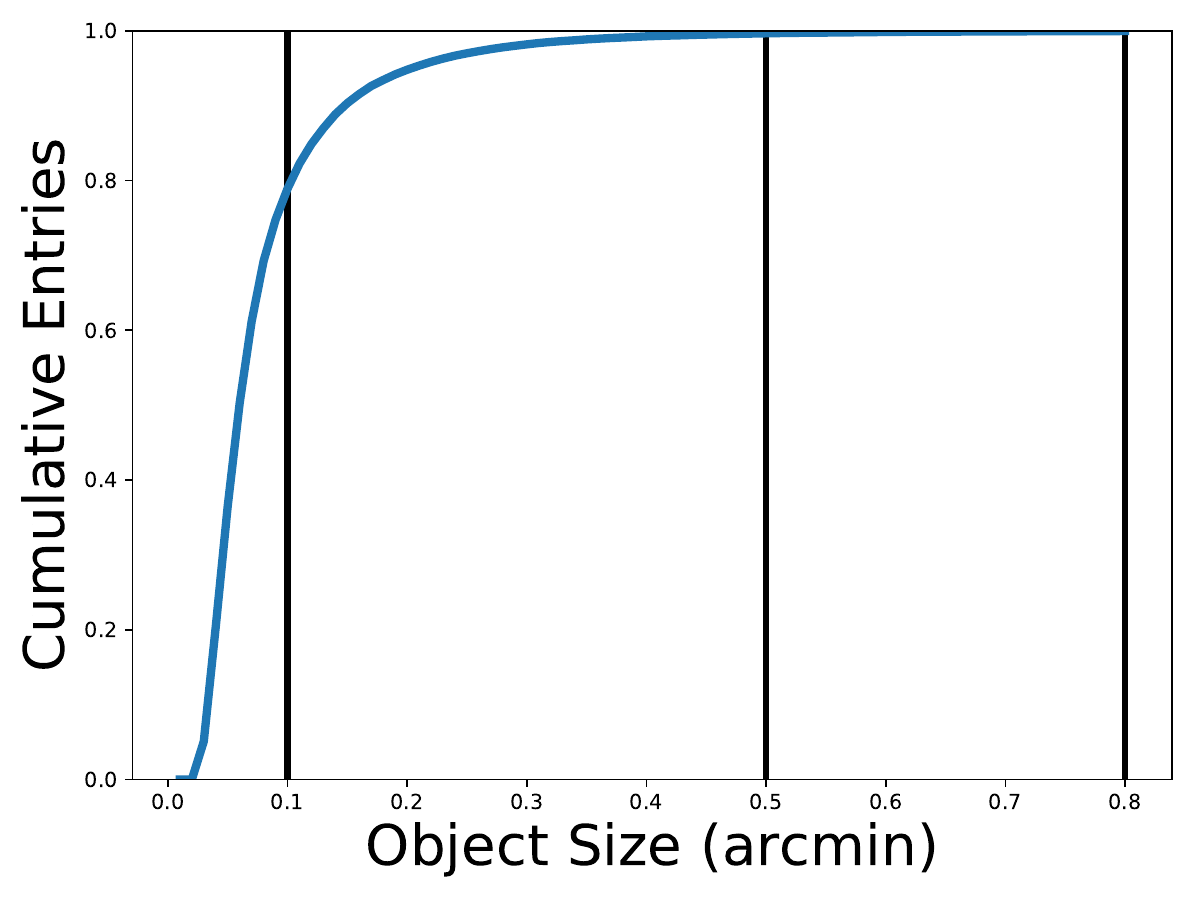}
\caption{
\edit{The cumulative distribution of \edit{blended galaxy} sizes from the HSC UltraDeep data with 
vertical lines indicating the cell sizes we chose.}
}
\label{fig:size}
\end{figure}

To connect the \edit{cell sizes} with observations, we need to know the typical 
object size in LSST.  Because we want to avoid splitting blended galaxies, we use the 
combined size of all galaxies identified as coming from the same blend. 
\edit{We can estimate these sizes from the HSC UltraDeep~\citep{HSC_DR1}}
Data Release 1 which has a depth of 
$\sim 27.5$, comparable to the wide field LSST depth. 
As a proxy for size, we use the square root of the area of the bounding box for all blended
objects detected in the HSC data. 
Fig.~\ref{fig:size} shows the cumulative distribution along with vertical lines for 
our chosen patch sizes.  It indicates that $\sim80\%$ of objects would fit in a patch size of 
0.1 arcminutes and we reach $\sim99\%$ at a size of 0.36 arcminutes.  We therefore conclude 
that a cell size near 1 arcminute will fully contain almost all objects.

\edit{\section{Building Cell-Based Coadds for LSST}\label{Section:Cell}}
\edit{Another challenge for edge-free coadds is how to choose a region of interest to coadd. 
For our simulations above, we used a per-object coadd. This presents challenges for real data
that needs to deal with the complications of object identification and galaxy blending \citep{Melchior2021}. 
A more practical approach is to build coadds in small regions on the sky called ``cells". 
Larger cells reject more exposures, but are less sensitive to the details of how galaxies are defined.}

\begin{figure*}
\includegraphics[width=\textwidth]{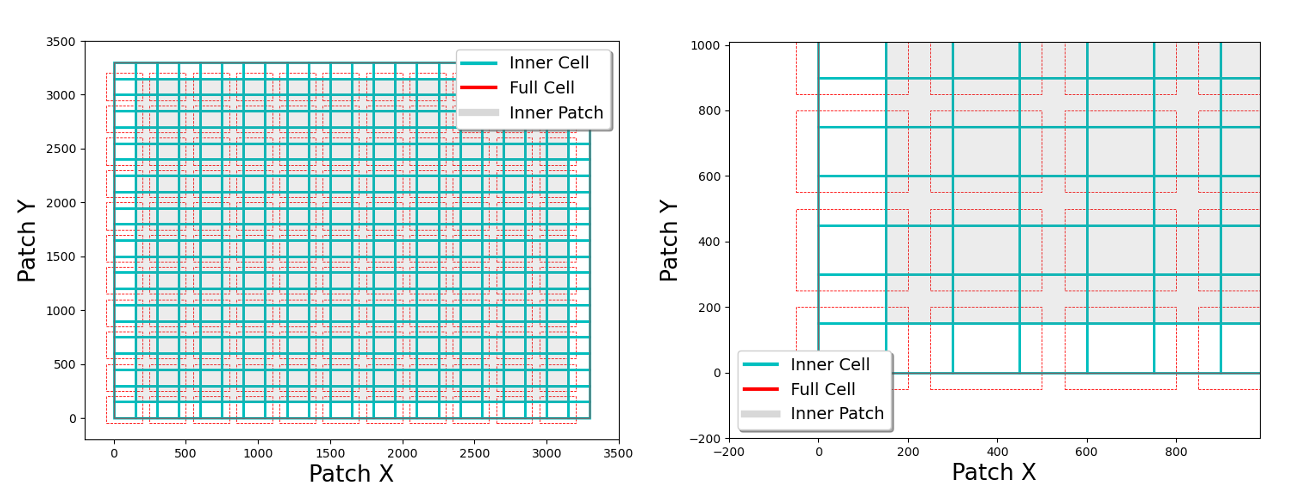}
\caption{
An example layout of cell-based coadds. {\it Left} : A full patch is displayed 
with the inner boundary of each cell outlined cyan. The full cell is outlined 
in red, but is only drawn for every other cell to reduce the number of lines on 
the plot. The grey region indicates the inner boundary of the patch.
{\it Right} : A zoomed in view of the patch near the lower left-hand corner, 
where again only every other full cell is plotted to reduce the number of lines.
}
\label{fig:cellsky}
\end{figure*}

\edit{To implement cell-based coadds for LSST data processing, we must add an additional 
layer of complexity to account for cells. Our approach for building cells is based on a strategy
devised for the Dark Energy Survey\citep{Becker2024} and adapted to the LSST pipeline.
The full sky is divided 
into a set of square ``tracts" that share the same tangent plane projection. 
Each tract contains a grid of square ``patches" that are typically tens of 
arcminutes on a side and have some amount of overlap.
Given that we want the size of our cells to be at arcminute scales, we 
need to further subdivide each patch into overlapping cells.
To meet this requirement, the patches are defined as integral multiples of cells, and tracts are defined as integral multiples of patches.
 Fig.~\ref{fig:cellsky} shows a visual 
example of a single patch divided into multiple cells.
While the exact 
sizes are yet to be determined, a tract is typically on the order of 1-2 degrees on each side. 
}

\edit{The inner region of the cells tile the sky.
To account for objects that may extend beyond the inner regions of a given cell, an extra boundary area is 
included in each cell that will overlap with neighboring cells. This buffer area can reduce problems 
that come from edge effects within a cell. To remove duplicate objects from these regions, catalogs must be 
trimmed to the unique area of each cell. This buffer area increases the amount of storage needed 
for cell-based coadds by a factor of a few over the standard patch-based storage.}

An implementation of cell-based coadd data structure is currently being incorporated into the LSST software\footnote{\url{https://github.com/lsst/cell_coadds}} 
and will likely be the default images on which shear will be measured.
This data structure can hold multiple realistic noise realizations that will be required for shear estimation algorithms.
Cell-based coadds may also have additional 
benefits over the standard patch-based coadds. For example, because cell-based coadds are small, 
the PSF on the coadd will have little spatial variation. Therefore, it may be sufficient to use the PSF 
at the center of the cell. This would make the PSF fast to compute compared to patch-based coadds which
requires warping the PSF from individual exposures and coadding them on the fly for each evaluation of the PSF. For data with many epochs, this operation dominates the compute time and makes it difficult 
to process patches in a reasonable amount of time. We leave the details of how to best construct 
cell-based coadds for LSST to future work

\section{Summary}
\label{Section:Summary}
We have shown that weak lensing shear can be reliably inferred using coadded
images that have a well-defined PSF.  We tested two state-of-the art shear 
techniques, BFD and \mcal, and in both cases we detected no additional bias due 
to coaddition.  We see minimal loss of information using non-optimal weighted mean 
coadds for the case of relatively small PSF size variations (of order ten percent).  
The two shear inference methods have quite different assumptions, lending support to 
the notion that the use of coadds may be possible for other methods as well.  \edit{We expect
that other shear measurement methods would also be able work on coadds as long as they can 
correctly account for the correlated noise that arises during the coaddition process.}  
We have explicitly assumed in this study that the coadded noise field is roughly constant over the
size of an object.  The scale of noise variation will depend on the distortion,
warping, etc.\ which could potentially change for different surveys.

These tests were done on simple inverse variance weighted coadds which are
relatively easy to construct.  However, there are some cases where this
approach may not be optimal.  For example, using such simple coadds will likely
somewhat degrade the ability to classify stars and galaxies, as the width of the
coadded PSF is necessarily larger than that of the best input images.  In this
case there is a clear benefit to constructing optimal coadds.  Exploring this in
more detail is beyond the scope of this work.

We did not address science cases beyond weak lensing shear inference and simple flux measurements.  
Other science cases may have different needs.  For example, it is likely that some form of multi-epoch 
fitting will be needed for effective detection and characterization of moving objects.

The cost of applying our approach to real data is that epochs which
cross CCD boundaries must be rejected from the coadd, in order to preserve the
continuity of the PSF.  \edit{We describe a potential strategy for LSST that builds cell-based coadds 
where the size of a cell is a few arcminutes on a side. Shear measurements, and potentially others, would
be performed on these cells. Using a combination of existing data and simulations, we showed that
for cells of $\sim1$ arcminute on a side most objects would fit inside this cell and that the
resulting number of exposures we would reject on average would be somewhere between 10-20$\%$. This
would lead to a small loss in depth ($\sim0.1$~mag), but would dramatically increase
the processing speed while reducing complexity compared to multi-epoch fitting, without
a loss in accuracy.
}

This work is especially relevant for surveys with many exposures like LSST.  
Multi-epoch fitting will dominate the compute resources for such a survey and therefore 
anything that can alleviate that will be extremely helpful.  We have shown that the 
computational cost for \edit{shear estimation} can be significantly reduced by using using coadded 
images instead of the more optimal multi-epoch fitting approach.

The weak lensing literature includes studies that take a wide variety of approaches to shear
inference in multi-epoch data.  In this work, we have sought to place the choice of approach on 
a firmer foundation (at least for Nyquist sampled ground-based images) by exploring issues of 
shear bias, statistical uncertainty, and the handling of objects near the boundaries of detectors 
in detail.  While some implementation details were deferred to future work, our results suggest that current and future ground-based weak 
lensing studies that adopt a principled coaddition method of the sort described in this work may be
well-justified in the use of coadds for shear inference.

\section*{Acknowledgments}

This paper has undergone joint internal review in the LSST Data Management team and the LSST Dark Energy Science Collaboration.  The primary authors would like to thank the internal reviewers Joshua Meyers, Arun Kannawadi, and Andr\'{e}s Plazas Malag\'{o}n for their comments.

Author contributions are as follows: R. Armstrong performed the shear tests of the BFD method, ran the \textsc{OpSim} simulations and was the primary author of the paper. E. Sheldon performed tests of the \Mcal\ method, contributed to testing and deriving the analytic $S/N$ loss function, and provided feedback on the document. E. Huff contributed to deriving the analytic $S/N$ loss function. Eli Rykoff wrote the cell-based patch-building code for LSST data management. J. Bosch provided many useful suggestions and discussions and feedback on the paper. R. Mandelbaum gave many useful suggestions and provided feedback on various drafts of the paper. A. Kannawadi contributed text on cell-based coadds for LSST and gave many helpful comments as a reviewer. M. Becker contributed the ideas on how to embed cell-coadds into the Rubin tract/patch coordinate system. P. Melchior was involved in many early discussions and gave feedback on the paper. R. Lupton helped to shape the scope of the project through many early discussions.  Y. Al-Sayyed participated in discussions on implementing cell-based coadds into the LSST software.

This  material  is  based  upon  work  supported  in  part  by  the  National  Science  Foundation  
through  Cooperative Agreement 1258333 managed by the Association of Universities for Research in 
Astronomy (AURA), and the Department of Energy under Contract  No.  DE-AC02-76SF00515 with the SLAC 
National Accelerator  Laboratory.      Additional  LSST funding  comes  from  private  donations,  
grants  to  universities,  and in-kind support from LSSTC Institutional Members. RM is supported by the US Department of Energy Cosmic Frontier program, grant DE-SC0010118.

The DESC acknowledges ongoing support from the Institut National de 
Physique Nucl\'eaire et de Physique des Particules in France; the 
Science \& Technology Facilities Council in the United Kingdom; and the
Department of Energy, the National Science Foundation, and the LSST 
Corporation in the United States.  DESC uses resources of the IN2P3 
Computing Center (CC-IN2P3--Lyon/Villeurbanne - France) funded by the 
Centre National de la Recherche Scientifique; the National Energy 
Research Scientific Computing Center, a DOE Office of Science User 
Facility supported by the Office of Science of the U.S.\ Department of
Energy under Contract No.\ DE-AC02-05CH11231; STFC DiRAC HPC Facilities, 
funded by UK BEIS National E-infrastructure capital grants; and the UK 
particle physics grid, supported by the GridPP Collaboration.  This 
work was performed in part under DOE Contract DE-AC02-76SF00515.

This work was performed under the auspices of the U.S. Department of Energy (DOE) by Lawrence Livermore National Laboratory (LLNL) under Contract DE-AC52-07NA27344, with IM release number LLNL-JRNL-866228. Funding for this work was provided as part of the DOE Office of Science, High Energy Physics cosmic frontier program. This document was prepared as an account of work sponsored by an agency of the United States government. Neither he United States government nor Lawrence Livermore National Security, LLC, nor any of their employees makes any warranty, expressed or implied, or assumes any legal liability or responsibility for the accuracy, completeness, or usefulness of any information, apparatus, product, or process disclosed, or represents that its use would not infringe privately owned rights. Reference herein to any specific commercial product, process, or service by trade name, trademark, manufacturer, or otherwise does not necessarily constitute or imply its endorsement, recommendation, or favoring by the United States government or Lawrence Livermore National Security, LLC. The views and opinions of authors expressed herein do not necessarily state or reflect those of the United States government or Lawrence Livermore National Security, LLC, and shall not be used for advertising or product endorsement purposes.

\section*{Data Availability}

 While the simulated data we used for this analysis was not saved, the code used to generate the underlying data will be shared on reasonable request.

\bibliographystyle{mnras}
\bibliography{references}

\appendix
\section{\texorpdfstring{$S/N$}{} Loss from Coaddition} 
\label{Section:FluxSNAppendix}
Here we estimate the additional uncertainty in the measured flux when coadding,
for the example of ``matched-filter'' photometry, where a linear fit is 
performed
for the amplitude of a model.
We consider an optimal estimator for a single unknown parameter that is
linear in the observables. Let the model be $A\boldsymbol{m}$, where $A$ is a
scalar amplitude and $\boldsymbol{m}$ is a normalized signal model, or
template. Then the log-likelihood for $A$ (assuming a Gaussian signal
likelihood) and some data vector $\boldsymbol{d}$ is

\begin{align}
\edit{\log L = - \frac{1}{2}(\boldsymbol{d} - A\boldsymbol{m})^T\: C^{-1} (\boldsymbol{d} - 
A\boldsymbol{m}) - \frac{1}{2} \det((2\pi)^k C )}
\end{align}
where $C$ is the noise covariance \edit{and $k$ the size of $\boldsymbol{m}$}.  
The optimal estimator $\hat{A}$ is the
value that maximizes this expression for $A$. With some algebra, it can be
shown that this value is:
\begin{align}
\hat{A} = \frac{\boldsymbol{m}^T C^{-1} \boldsymbol{d}}{\boldsymbol{m}^T C^{-1} 
\boldsymbol{m}}
\end{align}
and that the variance of $\hat{A}$ is
\begin{align}
{\rm var}\hat{A} = \frac{1}{{\boldsymbol{m}^T C^{-1} \boldsymbol{m}}}
\end{align}

In the case of photometry, $\boldsymbol{m}$ is the normalized profile of the
star or galaxy, $A$ is the measured flux, and $\boldsymbol{d}$ is the set of
pixels on which the measurement will be made.  The equations above are general,
but for simplicity in what follows, we assume the noise comes from a uniform
background, so there is no signal in the covariance.
For a set of $N$ images of the same sky, the data is the concatenation of the
pixel values in each epoch, i.e., $\boldsymbol{d} = \edit{\{d_1, d_2, ..., d_n \}}$.
This allows the template $\boldsymbol{m}$ to be the concatenation of the
templates appropriate for each epoch, if for instance the PSF varies from
exposure to exposure.

Now suppose we coadd the images using a straight mean, such that $\langle
\boldsymbol{d} \rangle = \frac{1}{N}\sum\limits_i d_i$, and the covariance
matrix is $C_\coadd = \frac{1}{N^2}\sum\limits_i C_i$.  The template
$\boldsymbol{m}_\coadd$ is then the mean $\boldsymbol{m}_\coadd = 
\frac{1}{N}\sum\limits_i m_i$
and the resulting operation is:

\begin{align}
\hat{A}_\coadd = \frac{\boldsymbol{m}_\coadd^T C_\coadd^{-1} \langle 
\boldsymbol{d} \rangle}{\boldsymbol{m}_\coadd^t 
C_\coadd^{-1}\boldsymbol{m}_\coadd} 
\end{align}
with estimator variance
\begin{align} \label{eq:exact_var_coadd}
{\rm var}\hat{A}_\coadd = \frac{1}{\boldsymbol{m}_\coadd^T 
C_\coadd^{-1}\boldsymbol{m}_\coadd},
\end{align}

where the indices in these expressions run over epochs. The multi-fitting
method, by contrast, would use the optimal estimator for each epoch:

\begin{align}
\hat{A}_{\rm multi} = \frac{\sum\limits_i \boldsymbol{m}_i^T 
C_i^{-1}\boldsymbol{d}_i}{\sum\limits_i \boldsymbol{m}_i^T 
C_i^{-1}\boldsymbol{m}_i}
\end{align}
with estimator variance
\begin{align}
\label{eq:exact_var_multi}
{\rm var}\hat{A}_{\rm multi} = \frac{1}{\sum\limits_i \boldsymbol{m}_i^T 
C_i^{-1}\boldsymbol{m}_i}.
\end{align}

These variance estimators predict, for the case of background-only noise, that 
the variance of the coadd estimator is generally greater than or equal to that of
the optimal estimator using the original images, assuming the templates are
accurate.  This is because the averaged signal will generally be less varied
than the set of input images, such \edit{that} $\boldsymbol{m}^T C^{-1} \boldsymbol{m}
\sim \sum \boldsymbol{m}^2$ is smaller for the coadd.
In order to gain better intuition for this increased variance, we depart from
generality in the data, and adopt a toy model for the signal.  First, we assume
that the noise in all images \edit{has the same statistical properties}, with standard deviation $\eta$, such
that the estimators become

\begin{align}
\hat{A}_{\rm multi} = \frac{\sum\limits_i \boldsymbol{m}_i^T 
\boldsymbol{d}}{\sum\limits_i \boldsymbol{m}_i^T \boldsymbol{m}_i}&, ~~
{\rm var}\hat{A}_{\rm multi} = \frac{\eta^2}{\sum\limits_i 
\boldsymbol{m}_i^T \boldsymbol{m}_i}\\
\hat{A}_\coadd = \frac{\boldsymbol{m}_\coadd^T \langle \boldsymbol{d} 
\rangle}{\boldsymbol{m}_\coadd^T \boldsymbol{m}_\coadd}&, ~~
{\rm var}\hat{A}_\coadd = \frac{\eta^2/N}{\boldsymbol{m}_\coadd^T 
\boldsymbol{m}_\coadd}
\end{align}

We now assume that the template for the coadd is related to that
in the original images by perturbations due to small variations
in a scale of the object $\sigma$:

\begin{align}
\boldsymbol{m}_{\rm i} &= \boldsymbol{m}_\coadd + \Delta \boldsymbol{m}_i \\
&\approx \boldsymbol{m}_\coadd + \frac{\partial 
\boldsymbol{m}_\coadd}{\partial \sigma} \Delta \sigma
\end{align}

The ratio of variances then becomes

\begin{align} \label{eq:generalperturb}
\frac{ {\rm var} \hat{A}_\coadd }{{\rm var} \hat{A}_{\rm multi} } &=  
1 + \frac{1}{N}\frac{\sum\limits_i \Delta \model_i^T \Delta \model_i 
}{\modelc^T \modelc} \\
& \approx 
1 + \frac{(\Delta \sigma)^2}{N}\frac{\sum\limits_i \left( \frac{\partial 
\modelc}{\partial \sigma} \right)_i^T \left( \frac{\partial \modelc}{\partial 
\sigma} \right)_i }{\modelc^T \modelc}
\end{align}

We further assume the template is a round Gaussian

\begin{align}
\modelc = \frac{1}{2 \pi \sigma^2} e^{-r^2/2 \sigma^2 }
\end{align}
with derivative
\begin{align}
\frac{\partial \modelc}{\partial \sigma} &= \frac{1}{2 \pi \sigma^2} 
\frac{2}{\sigma} e^{r^2/2 \sigma^2} \Bigl( \frac{1}{2}\frac{r^2}{\sigma^2} - 1 
\Bigr).
\end{align}

For these Gaussian models, all terms in equation~\eqref{eq:generalperturb} can
be readily calculated in the continuous limit, and we find

\begin{align} \label{eq:varbasic}
\left( \frac{ {\rm var} \hat{A}_\coadd }{{\rm var} \hat{A}_{\rm multi} } 
\right)_{\rm toy} = 
1 + 2 \left( \frac{\Delta \sigma}{\sigma} \right)^2
\end{align}
where $(\Delta \sigma)^2$ is to be interpreted as the rms variation of the size.
We expect the increase in variance to be less for the case
of large galaxies convolved by a PSF.  If the galaxy is a round Gaussian with
size $\sigma_g$ and the PSF is a round Gaussian with size $\sigma_p$,
and only the PSF size varies between images,
we can use the chain rule to rewrite equation~\eqref{eq:varbasic} as

\begin{align} \label{eq:vargal_appendix}
\left( \frac{ {\rm var} \hat{A}_\coadd }{{\rm var} \hat{A}_{\rm multi} } 
\right)_{\rm toy} &= 
1 + 2 \left( \frac{\sigma_p^2}{\sigma_g^2 + \sigma_p^2}\right)^2 \left( 
\frac{\Delta \sigma_p}{\sigma_p} \right)^2 \\
&\equiv 1 + 2 \left( 1 - R \right)^2 \left( \frac{\Delta 
\sigma_p}{\sigma_p} \right)^2
\end{align}
where we have used the definition of the resolution factor $R =
\sigma_g^2/(\sigma_p^2 + \sigma_g^2)$.  This confirms our
intuition that measurements of large galaxies, with $R \sim 1$ will suffer
less increase in variance.
Note that, for this toy model, the template is not exactly equal to the mean of
the individual templates.  Thus we expect the toy model to slightly
over-predict the increase in variance due to coaddition.

\end{document}